\documentclass[twocolumn,amsmath,amssymb,aps]{revtex4}
  \usepackage[dvips]{graphicx}
\begin{document}

\title{Stochastic variability of oceanic flows above topography anomalies}

\author{Antoine Venaille}

 \altaffiliation[Current Affiliation: ]{Phys. ENS-Lyon, 69007 Lyon France}
 \email{antoine.venaille@ens-lyon.org}
\author{Julien Le Sommer}
\author{Jean-Marc Molines}
\author{Bernard Barnier}
\affiliation{LEGI/UJF/CNRS/INPG, Grenoble, France}

\date{\today}

\begin{abstract}
We describe a stochastic variability mechanism which is genuinely  internal to the ocean, i.e. not due to fluctuations in atmospheric forcing. 
The  key ingredient is the existence of closed contours of bottom topography surrounded by a stirring region of enhanced eddy activity. 
This configuration leads to the formation of a robust but highly variable vortex above the topography anomaly.
The vortex dynamics integrates the white noise forcing of oceanic eddies into a red noise signal for the large scale volume transport of the vortex.
The strong interannual fluctuations of the transport of the Zapiola anticyclone ($\sim 100 \ Sv$) in the Argentine basin are argued to be partly due to such eddy-driven stochastic variability, on the basis of a $310$ years long simulation of a comprehensive global ocean model run driven by a repeated-year forcing.
\end{abstract}

\maketitle

\section{Introduction}

Large-scale oceanic currents show large fluctuations at decadal, centennial and even millennial time scales \citep{Ghil02}. 
Understanding the nature and the causes of this variability is a major challenge for climate predictions. 
Some properties  of oceanic  variability can result from the intrinsic, deterministic non-linear dynamics of oceanic flows \citep{DijkstraGhil05}. 
For instance, this approach has been proven useful to interpret the different peaks arising  at decadal time scales in time-series  of the Gulf-Stream position \citep{Simonnet05}. 
Otherwise, the ubiquity of red noise signals in time series of various oceanic metrics is generally thought to result from the integration of white noise processes by the linearized dynamics \citep{Hasselman76}. 
Obviously, a source of noise is associated with the  rapid fluctuations of atmospheric forcing, therefore  leading to low frequency oceanic variability \citep{Frankignoul97}.
However, red noise signals have also been reported in  ocean models forced by steady winds but with sufficient spatial resolution to simulate the ``oceanic weather system'', i.e. mesoscale eddies at scale from $30$ to $300$ km \citep{BerloffMcWilliams,Hogg05}.  This suggests the existence of intrinsic, stochastic variability of large-scale oceanic currents. 

We show in the following that the dynamics of eddy-driven flows above closed topography contours provides essential ingredients for supporting such stochastic variability  genuinely  internal to the ocean. We then argue that this mechanism contributes to the variability of the Zapiola anticyclone in the Argentine Basin. 

\section{Basic mechanism}

\begin{figure}
\begin{center}
 \noindent\includegraphics[width=0.5\textwidth]{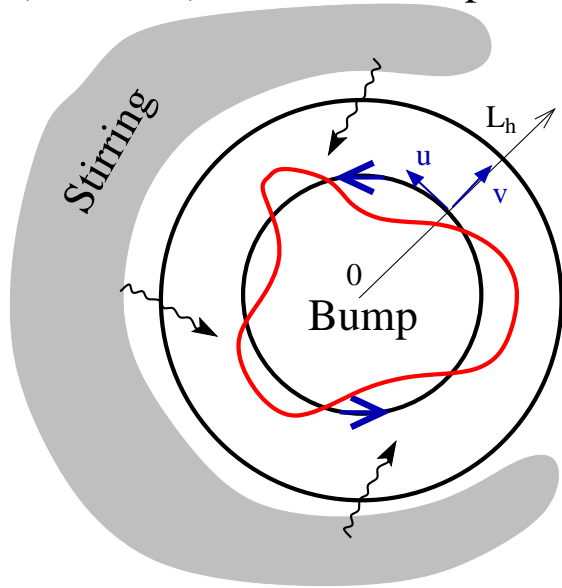}
\end{center}
\caption{ \textbf{Large scale flows driven by eddies in presence of a mean gradient of potential vorticity $f/h$.} \textbf{a)} Case of mid-latitude atmospheric circulation: the potential vorticity gradient is due to variations of the Coriolis parameter $f$ with latitude (beta effect). Arrows represent Rossby waves that propagate away from the stirring region and imply the formation of a westward flow outside the stirring region and an eastward flow in the stirring region 
\cite[][adapted from figures 12.2 and 12.3]{VallisBook}.  \textbf{b)} Case of oceanic flows above a topographic bump of typical length $L_h=600$  km: the potential vorticity gradient is due to variations of depth $h$.}
\label{fig:schema}
\end{figure}

Let us consider first a barotropic (depth independent) flow above an axisymmetric topographic bump $h(r)$ such that the total fluid depth $h(r)$ is increasing  for $r<L_h$, and constant for $r>L_h$, and surrounded by a ``stirring'' region, as sketched on figure \ref{fig:schema}-b. 
Here, stirring can be due to any mechanism outside the topography anomaly, generating disturbances in the flow field that propagate toward the topography anomaly. This situation is reminiscent of the large scale atmospheric circulation induced by mid-latitude stirring \citep{williams}. 
Simple arguments relying on Kelvin circulation theorem  explain the formation of an anticyclonic flow above the closed topographic contours\footnote{{Strictly speaking, one should consider $f/h$ contours rather than topography contours}} in the same way as westward flows are formed away from stirring regions in the atmosphere, see e.g. \citep{Held75,VallisBook}. 
%
%Note that contrary to the atmospheric case, the stirring region is here located outside the topography anomaly. Therefore, the formation of a cyclonic flow pattern in the stirring region,  equivalent to the mid-latitude eastward jet in the atmosphere, is not expected in this case.
%
Neglecting other forcing than stirring, and considering that dissipation occurs through bottom friction modeled as a linear drag with coefficient $\omega_b$, the dynamics above the topography anomaly reads 

\begin{equation}
\partial_t q +\mathbf{u} \cdot \nabla q=  -\omega_b \frac{\zeta}{h} \ , \quad \nabla \cdot \left(h \mathbf{u} \right) =0 \ , \quad q=\frac{\zeta+f}{h}\ ,\label{eq:PVdyn}
\end{equation}

where $\mathbf{u}$ is the velocity,  $\zeta=(\nabla\times \mathbf{u}) \cdot \mathbf{e}_z$ is the relative vorticity and $f$ is the planetary vorticity, taken here as a constant {(so that  $f/h$ contours are also topography contours)}.
$\mathbf{e}_z$ is a unit vector pointing upward. 

We call $v$ and $u$ the radial and azimuthal velocities respectively ($v$ is positive when directed outside the topography anomaly,  $u$ is positive when anti-clockwise). 
Defining eddies as deviation from the azimuthal average $u^\prime=u-\overline{u}$ and integrating equation (\ref{eq:PVdyn}) inside a circle of constant $r$ yields 
\begin{equation}
\frac{\partial}{\partial t} \overline{u} = -\frac{1}{r^2}\frac{\partial}{\partial r}  \left(r^2 \overline{u^\prime v^\prime} \right) - \omega_b \overline{u} .
\label{eq:u}
\end{equation}
Multiplying equation (\ref{eq:u}) by the total fluid depth  $h(r)$ and integrating from the center $r=0$ to the boundary $r=L_h$ of the topography anomaly yields the  transport equation 
\begin{equation}
\frac{d {\mathcal{T}}}{dt} =  \eta_{eddy} - \omega_b {\mathcal{T}}, 
\label{eq:AR1}
\end{equation}
where $\mathcal{T}=  \int_{0}^{L_h}\  \mathrm{d}r \ h \overline{u}$. Assuming that  typical variations of $h$ over $L_h$ are smaller than $h$, and assuming that $ |\overline{u'v'}/r| \ll |\partial_r \overline{u'v'}|$ (which corresponds to the limit case of a channel), the eddy term $\eta_{eddy} \approx  -h \,\overline{v^\prime u^\prime}|_{r=L_{h}}$ is given by the eddy momentum fluxes at the boundary of the bump.
%\begin{equation}
%\eta_{eddy}=  -h \,\overline{v^\prime u^\prime}|_{r=L_{h}} + \int_{0}^{L_h} \mathrm{d}r \   {\overline{v^\prime u^\prime}} r^2 \frac{\partial}{\partial r} \left( \frac{h}{r^2}\right) \ . 
%\label{eq:AR1bis}
%\end{equation}
%provided that typical variations of $h$ over $L_h$ are smaller than $h$, and that $ |\overline{u'v'}/r| \ll |\partial_r \overline{u'v'}| $ (which corresponds to the limit case of a channel), the second term of the right hand side of (\ref{eq:AR1bis}) can be neglected, and $\eta_{eddy} \approx  -h \,\overline{v^\prime u^\prime}|_{r=L_{h}}$ is then given by the eddy momentum fluxes at the boundary of the bump. The latter expression will be used in the following for diagnosing $\eta_{eddy}$ with model output.
%where $\eta_{eddy}(t)$ is the vertically and azimuthally integrated eddy momentum flux through the boundary $r=L_h$.

{In the absence of direct external forcing, eddy fluxes tend to reduce large scale potential vorticity gradients, therefore driving anticyclonic flow above topographic bumps \citep{BrethertonHaidvogel}.
 One expects then that time-mean of the eddy term $\eta_{eddy} $ is positive (resp. negative) in the case of a positive (resp. negative) topography anomaly, as for instance predicted by linear theory \citep{Held75}, and as reported in various numerical simulations, see e.g. \citep{Holloway,Treguier}.} 
If $\eta_{eddy}$ were time independent,  the vortex would reach a stationary state and its transport would be inversely proportional to bottom friction \citep{Dewar98}. However, stirring is more generally time varying, and so is the driving term $\eta_{eddy}$. In particular, the eddy term  $\eta_{eddies}$  could possibly vary depending on the anticyclone transport $\mathcal{T}$, thus allowing complex feedbacks as in the atmosphere \citep{LorenzHartmann01}.
Assuming here that such feedbacks are negligible, and modelling the eddy term $\eta_{eddy}(t)$ as a constant plus a gaussian white noise, the transport satisfies a Langevin equation \citep{Langevin}. 
Transport time series are then instances of an auto-regressive process of order one  \citep{Hasselman76,VonStorch}. 
For frequency $\omega$ higher than  the  cut-off frequency $\omega_b$ set by bottom friction, the transport power spectrum  $|\widetilde{\mathcal{T}}(\omega)|^2 =  |\widetilde{\eta}_{eddy}|^2 (\omega^2+\omega_b^{2})^{-1}$ is a red noise. 
This shows that  oceanic eddies can drive the formation and the low frequency variability of anticyclonic flows above topographic bumps.

We argue in the following that this mechanism is responsible for the massive fluctuations in transport time series of the Zapiola Anticyclone \citep{Saraceno09}. 

\section{Intrinsic variability of the Zapiola anticyclone}

\begin{figure*}
\begin{center}
\noindent\includegraphics[width=\textwidth]{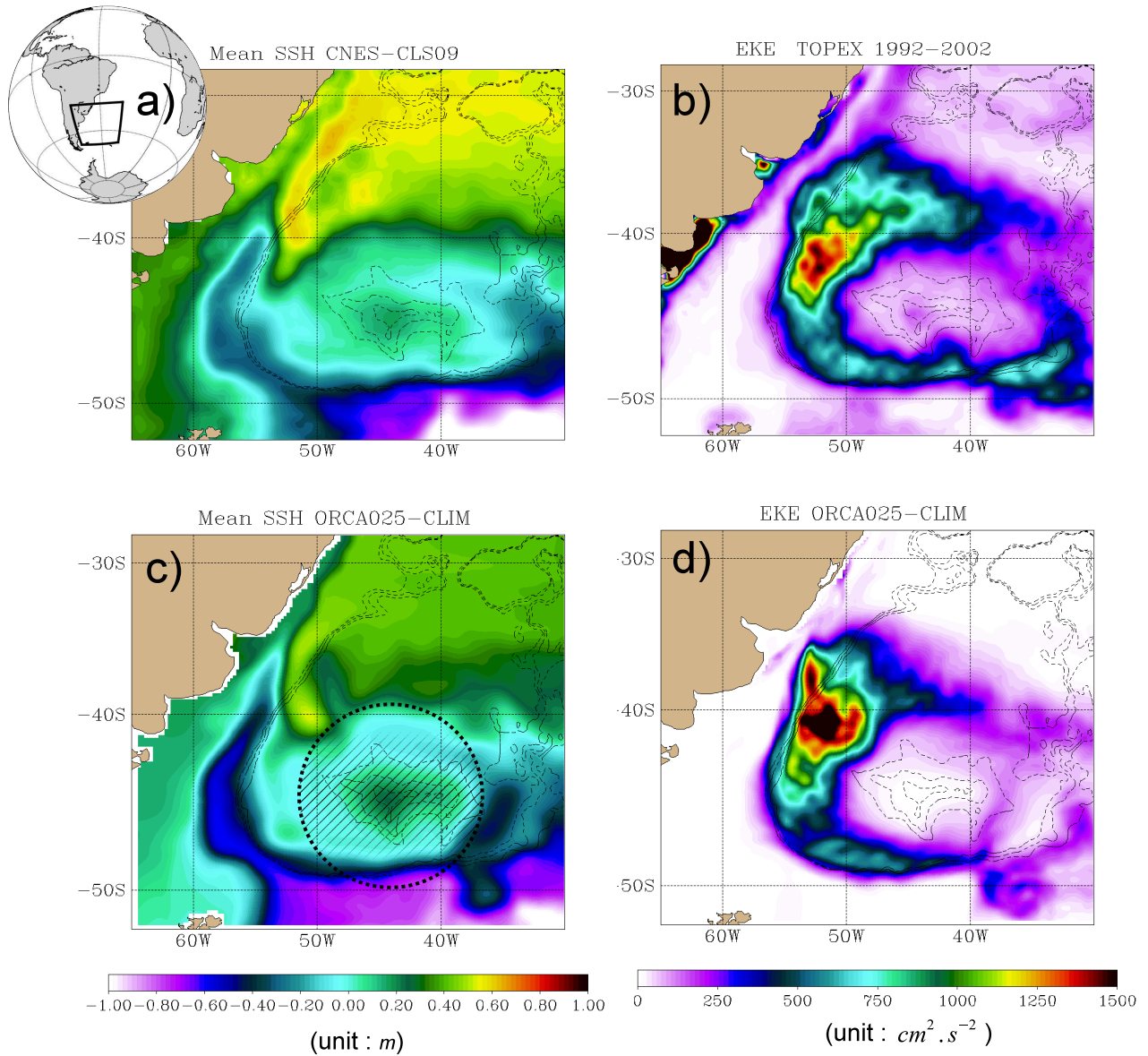}
\end{center}
\caption{ \textbf{The Zapiola anticyclone in the Argentine Basin.} \textbf{a)} Mean Dynamical Topography (SSH)  from CNES-CLS09 dataset. 
Above the Zapiola drift, SSH is proportional (and opposite) to the transport streamfunction since the flow is mostly depth independent.
\textbf{b)}  Surface eddy kinetic energy estimated from TOPEX altimeter over 1992-2002. 
\textbf{c,d)} Same quantities as a,b) in the model output over the last 10 years of the simulation. Isolines of $f/h$, where $h$ is the local depth, are represented in thin dashed black line.
The shaded circle in the lower left panel shows the region over which the diagnostics have been performed.}
\label{fig:ZapiolaMean}
\end{figure*}

The Zapiola anticyclone is a strong and robust anticyclonic vortex above a sedimentary bump  (the Zapiola drift), located in the Argentine basin, around $(45^{\circ}S,~45^{\circ}W)$, see figure \ref{fig:ZapiolaMean}.
This bump is characterized by shoaling  bottom topography rising by $\sim 800$ m  over a distance of $\sim 400$ km. 
The  anticyclone appears as a local maximum in sea surface height as shown in  figure \ref{fig:ZapiolaMean}-a. 
Typical velocities of the anticyclone are about $0.1\ m.s^{-1}$ over the whole vertical water column ($\sim 5000$ m depth). 
The associated transport is of the order of $100 \  Sv$ ($ = 10^6 \  m^3.s^{-1}$) which is comparable with other major oceanic currents including the Gulf-Stream. 
Remarkably, the Zapiola anticyclone is surrounded by one of the most eddying region of the global ocean, namely the confluence between the Nord-Brazil current and the Malvinas current (see figure \ref{fig:ZapiolaMean}-b). 
Perhaps surprisingly, the Zapiola anticyclone has only been discovered rather lately, and almost simultaneously with in situ measurement \citep{Weatherly93,SaundersKing95},   numerical ocean models \citep{deMiranda99}, and in theoretical studies \citep{Dewar98}. 
Also striking is the strong interannual variability of the Zapiola anticyclone transport which has recently been reported from  $15$ years of altimetric measurements \citep{Saraceno09}. 
High resolution data syntheses also suggest the dominance of interior ocean dynamics as opposed to atmospheric forcing in determining the variability of the Zapiola anticyclone  \citep{VolkovFu08}. 
But the strong variability of the Zapiola anticyclone transport remains unexplained.

Here, we report strong  interannual fluctuations of the Zapiola anticyclone transport in a 310 year long simulation of the global eddy admitting DRAKKAR model configuration \citep{barnier06,Drakkar2007} under a repeated-year forcing. 
This model configuration is based on the NEMO ocean/sea-ice modeling framework \citep{Madec2008}. The model grid uses $z$-coordinates with 46 levels and a partial step representation of bottom topography. The model horizontal resolution at the equator is 1/4th degree and varies with the cosine of the latitude so that at 45$^{\circ}$S the horizontal resolution is about $20km$. The model configuration uses a biharmonic viscosity for momentum and an isoneutral harmonic diffusivity for tracers. The bottom friction is a quadratic function of the bottom velocity $\mathbf{u_b}$ so that the bottom stress $\mathbf{\tau}_b$ is given by
$$
\mathbf{\tau}_b=C_d \left(\mathbf{u_b}^2+e \right)^{1/2} \mathbf{u_b},
$$ 
where $e=2.5\; 10^{-3}$ m$^2$.s$^{-2}$ is the bottom turbulent kinetic energy due to tides and other small scale processes. According to common practice, the coefficient $C_d$ is calibrated such that for weak barotropic flows, the bottom friction time-scale $\omega_b^{-1}$ is about $30$ months. For a barotropic flow of about $0.1$m.s$^{-1}$ as the Zapiola anticyclone,  the bottom friction time-scale $\omega_b^{-1}$ is about $14$ months.

The repeated-year forcing is built from the DRAKKAR forcing set DFS4 \citep{Brodeau2009}. The forcing was built by computing a daily climatology of the variables available in DFS4 during 50 years at 6-hourly periods. Notably, the quadratic quantities (wind stress and bulk exchange coefficients) were also averaged in order to guarantee that the climatological forcing is energetically consistent \citep{Penduff11}. The resulting forcing fields were then low-passed filtered in time (3-point hanning filter) to remove the remaining high-frequency noise. The resulting one year long  forcing was applied repeatedly during the 310 years of the simulation.

\begin{figure*}
\begin{center}
\noindent\includegraphics[width=0.9\textwidth]{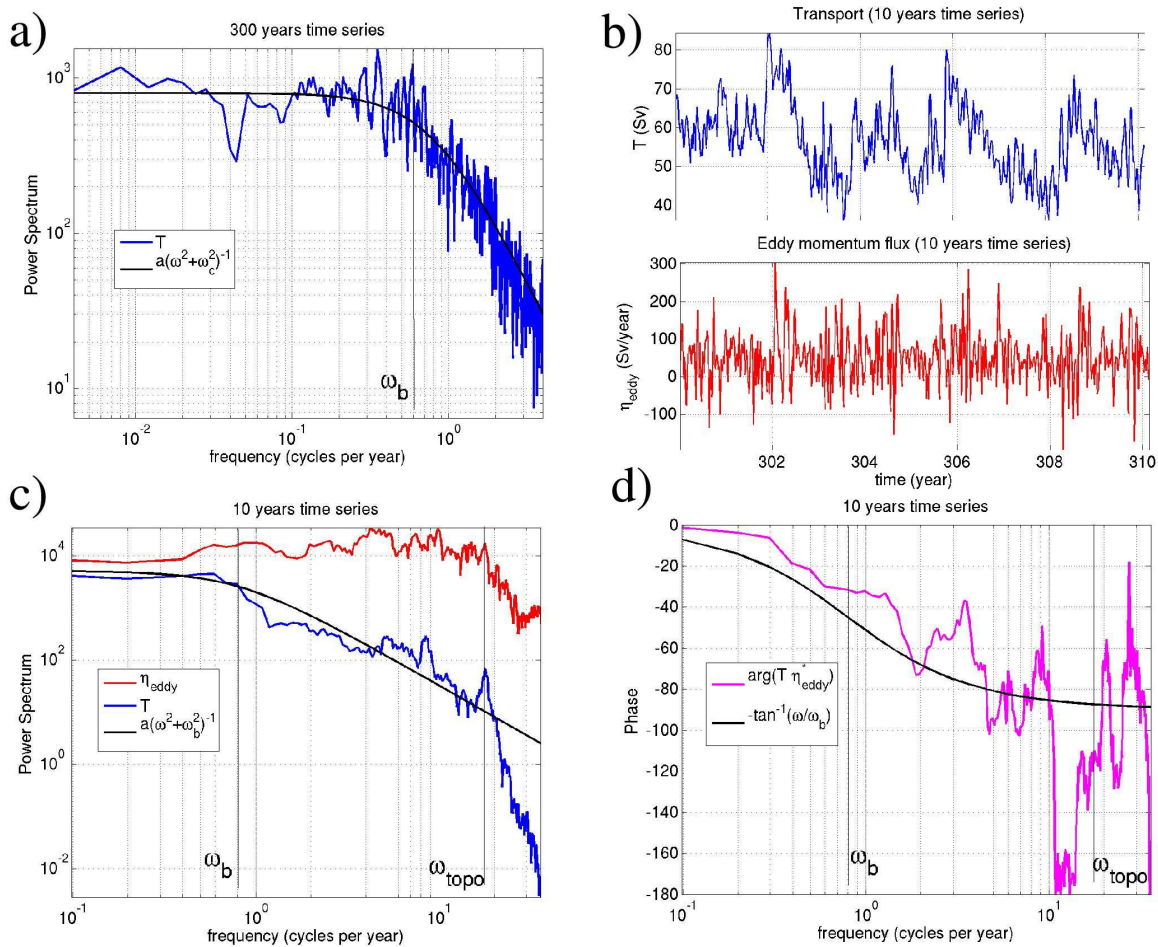}
\end{center}
\caption{ \textbf{Transport fluctuations.} \textbf{a)} Transport power-spectrum {$|\widetilde{\mathcal{T}}|^2(\omega)$} from the 300 years time series of monthly mean model output. \textbf{b)} Transport and eddy momentum flux time series from the last 10 years time serie of 5 days mean model output \textbf{c)} Transport  and eddy momentum flux  power-spectrum from the 10 years time series of 5 days mean model output.  \textbf{d)} {Phase $\phi$ of the cospectrum $\widetilde{\mathcal{T}}\widetilde{\eta}_{eddy}^*$}.}
\label{fig:timeseries}
\end{figure*}

Figure  \ref{fig:timeseries}-a,b  show strong fluctuations of transport over the last 10 years of the simulation and the associated power-spectrum computed over the first 300 years of the simulation. The model being forced by repeated-year atmospheric fields, the wind forcing is practically identical every year. Therefore, in the model,  interannual fluctuations of Zapiola anticyclone transport can only be due to intrinsic ocean dynamics. This experiment supports the idea that the variability of the Zapiola anticyclone can be partly attributed to intrinsic ocean dynamics.

A comparison of the time-series of transport $\mathcal{T}$ and the time-series of the eddy momentum flux $\eta_{eddy}$ is performed over the last  $10$ years of the simulation,  see figure \ref{fig:timeseries}. 
Remarkably, fluctuations of the eddy momentum flux are essentially white at frequency $\omega$ lower than $\omega_{topo} \sim$ ($20$ days)$^{-1}$ (figure \ref{fig:timeseries}-c).  
This correlation time scale corresponds to the period of a barotropic dipole rotating around the Zapiola drift \citep{Fu01,HughesJGR07}, and interpreted either as a topographic Rossby wave  \citep{Fu01}, or as the superposition of Rossby basin modes \citep{WeijerJPOa}. 
As expected from equation (\ref{eq:AR1}), for frequency $\omega$ lower than $\omega_{topo}$,  the transport power spectra of figures  \ref{fig:timeseries}-a and \ref{fig:timeseries}-c are well fitted by the function $a (\omega^2+\omega_b^2)^{-1} $, where the cut-off frequency $\omega_b$ is estimated by evaluating the ratio between time-mean transport and time-mean eddy momentum flux $\left<\eta_{eddy}\right>/\left<T\right> \sim$ ($15$ months)$^{-1}$, and where the coefficient $a$ is the low-frequency asymptotic value of the transport power spectrum.  
Strikingly, the value of the cut-off frequency $\omega_b$ estimated from the transport and eddy momentum flux time-series is of similar magnitude as the effective bottom friction coefficient used in the model (estimated at about $(14$ month$)^{-1}$, see appendix).
{According to equation (\ref{eq:AR1}), the phase of the cospectrum $\widetilde{\mathcal{T}} \widetilde{\eta}_{eddy}^*$ should be given by $-\tan^{-1}(\omega/\omega_b)$. 
This prediction is again consistent with model output presented figure \ref{fig:timeseries}-d, despite some discrepancy for frequency of about month$^{-1}$.
The discrepancy could come either from baroclinic effects or from possible retroaction of the mean flow on the eddies.}
The above diagnostics support the idea that the red noise spectrum of the Zapiola anticyclone transport time-series (at seasonal to interannual time scale, see figure  \ref{fig:timeseries}-a) results from the integration  by the dynamics of the eddy momentum flux fluctuations at the boundary of the anticyclone. 
{Note that in presence of wind fluctuations, this integration mechanism would account for an additional low frequency variability of the anticyclone transport, just as in \cite{WeijerGilleJPO05}.}

\section{Conclusion}

In this letter, we have described a simple mechanism contributing to low-frequency variability of oceanic flows above large scale (around $500$ km) topography anomalies.  
The key ingredients are closed contours of topography and a source of eddies. 
Our diagnostics of a comprehensive numerical ocean model suggest that such a mechanism explains at least part of the variability of the Zapiola Anticyclone in the Argentine Basin.
%
%The model configuration was such that this variability is necessarily intrinsic to the ocean dynamics.
%
%However, in presence of variable  wind forcing, the same mechanism could also explain the integration of such external wind fluctuations into low-frequency variability in the transport of the anticyclone. 
%

The formation of robust eddy-driven vortices above topography anomalies may provide an effective route for energy dissipation in the ocean. 
Indeed, because of the unstable nature of large scale oceanic flows, mesoscale eddies are ubiquitous in the ocean.
In particular, high values of mesoscale eddy kinetic energy are observed along western boundaries of ocean basins. 
Here,  we have depicted a situation where the mesoscale eddy kinetic energy can be absorbed above some topography anomaly (the Zapiola drift) and transferred to mean kinetic energy (the Zapiola anticyclone). 
The energy transfer is associated with the excitation, and the propagation of topographic Rossby waves \citep{FuJPO07}. 
%
%\textbf{Note that another source of eddy kinetic energy could come from Rossby waves genetared and the east side of the anticycline \cite{MarshallJMR}}
%
A simple estimation of the order of magnitude for this dissipation rate in the case of the Zapiola anticyclone gives a value of about $1.\ 10^{-3}$ W.m$^{-2}$ over an area $A \sim 1.\  10^{6}$ km$^2$, which amounts to a total of $1. 10^{-3}$ TW. 
This amounts to a significant fraction (about one tenth) of the total sink of ocean-eddy energy at the western boundary of the South Atlantic  \citep{MarshallGeoscience}.
Given the simple ingredients involved in setting the eddy-driven flow presented above, one might conjecture that the mechanism described in this letter could play an significant role in the internal low frequency dynamics and energetics of large scale oceanic currents.

%%

%%% End of body of article:

%%%%%%%%%%%%%%%%%%%%%%%%%%%%%%%%
%% Optional Appendix goes here
%
%%%%%%%%%%%%%%%%%
% Geophysical Research Letters only allows an appendix without a letter.
%% You can get this result with
%  \section*{Appendix}
%  or
 \section*{Appendix: diagnostic details}

The barotropic transport streamfunction $\psi$ is defined according to  $h\mathbf{u}= \mathbf{e}_z \times \nabla \psi $, where $\mathbf{u}$ is the depth averaged velocity. The anticyclone transport is defined as $\mathcal{T}=\overline{\psi}(boundary) -\psi(center)$, where $\psi(center)$ is the value of the streamfunction above the Zapiola drift ($45E45S$), and  $\overline{\psi}(boundary)$ is the average of the streamfunction along the circle of radius  $L_h \sim 500$ $km$ depicted in figure \ref{fig:ZapiolaMean}-c. 

The term $\eta_{eddy}$ is estimated by computing the term $\overline{h} \overline{u v}$, with azimuthal averages taken along the same circle $r=L_h$. Output data of the 310 years simulation are monthly averaged, except for the last ten years of the simulation, where 5-days averages  have been computed. Eddy terms can therefore only be computed over the last 10 years.

\begin{acknowledgments}
The model simulation was performed at CINES (Montpellier, France) as part of the "Grands Challenges GENCI/CINES 2008" and benefited from the technical expertise of Nicole Audiffren. JLS, JMM and BB are supported by the CNRS. AV was supported by DoE grant DE-SC0005189 and NOAA grant NA08OAR4320752 during part of this work. We thank Isaac Held for suggesting the analogy with the atmosphere, as well as David Marshall and  Wilbert Weijer for their constructive review. The diagnostics presented in this letter are inspired from a poster presented by Chris W. Hughes at the European Geophysical Union General Assembly 2008 under reference EGU2008-A-06866.
\end{acknowledgments}

\end{document}